\documentclass[12pt]{article}
\usepackage[english]{babel}
\usepackage{epsfig}
\begin{document}
\large

\par
\noindent {\bf Importance of the Mechanism of Resonance
Enhancement of Neutrino Oscillations  in Matter for the Precise
Testing of the Electroweak Interaction Model. Present Experimental
Status of This Resonance Mechanism.}
\par
\begin{center}
\vspace{0.3cm} Beshtoev Kh. M. (beshtoev@cv.jinr.ru)
\par
\vspace{0.3cm} Joint Institute for Nuclear Research, Joliot Curie
6, 141980 Dubna, Moscow region, Russia.
\end{center}
\vspace{0.3cm}

\par
Abstract

\par
The mechanism of resonance enhancement of neutrino oscillations in
matter and some critical remarks to this mechanism are considered.
Using of this resonance mechanism is very important to examine the
model of electroweak interactions since the processes induced by
this mechanism grow multiply. In contrast to the electromagnetic
and strong interactions in weak interactions, $P$-parity is
violated therefore a problem of mass generations in the weak
interactions is considered (the interaction must be left-right
symmetric for mass generations). It is concluded that a
possibility of mass generation in the framework of the weak
interactions is not proved.

The present experimental status of this resonance mechanism is
considered and it is done conclusion that this effect has no clear
experimental confirmation. For this purpose it is necessary to
fulfil precision experiments with solar neutrinos and the
neutrinos passed through the Earth matter.    \\



\newpage
\section{Introduction}

\par
The suggestion that, in analogy with $K^{o},\bar K^{o}$
oscillations, there could be neutrino-antineutrino oscillations (
$\nu \rightarrow \bar \nu$), was made by Pontecorvo \cite{1} in
1957. It was subsequently considered by Maki et al. \cite{2} and
Pontecorvo \cite{3} that there could be mixings (and oscillations)
of neutrinos of different flavors (i.e., $\nu_{e} \rightarrow
\nu_{\mu}$ transitions).
\par
The first experiment \cite{4} on the solar neutrinos has shown
that there is a deficit of neutrinos, i.e., the solar neutrinos
flux detected in the experiment was few times smaller than the
flux computed in the framework of the Sun Standard Model \cite{5}.
The subsequent experiments and theoretical computation have
confirmed the deficit of the solar neutrinos \cite{6}.
\par
The short base reactor and accelerator experiments \cite{7} have
shown that there is no neutrino deficit, then this result was
interpreted as an indication that neutrino vacuum angle mixing is
very small. Then the question arises: what is the deficit of the
solar neutrinos related? In 1978 the work by L. Wolfenstein
\cite{8} appeared where an equation describing neutrino passing
through the matter was formulated (afterwards that equation was
named Wolfenstein's). In the framework of this equation the
enhancement of neutrino oscillations in matter arises via weak
interactions. This mechanism of neutrino oscillations enhancement
in the matter attracted attention of neutrino physicists after
publications \cite{9} by S. Mikheyev and A. Smirnov where it was
shown that in the framework of this equation the resonance
enhancement of neutrino oscillations in matter will take place.
Also it is clear that there adiabatic neutrino transitions can
arise in matter if effective masses of neutrinos change in matter
\cite{10}. After that an enormous number of works appeared where
the deficit of the solar neutrinos was explained by this
mechanism. It is supposed that neutrino vacuum angle mixing is
very small \cite{11 totsuka} and at resonance enhancement of
neutrino oscillations in the solar matter this angle becomes
maximal ($\pi/4$). This mechanism was recognized as the only
mechanism to explain the origin of the Sun neutrino deficit and it
is supposed that the vacuum angle mixing is very small. The
situation changes after detection that the atmospheric neutrinos
angle mixing \cite{12 s-k atm} is big and close to the maximal one
$\pi/4$. The $ \bar \nu_e \to \bar \nu_{\mu}$ angle mixing
obtained in KAMLand detector \cite{13 KamLand} appears to be big
and near to the maximal one. Then the Day-Night effect does not
obtain a confirmation \cite{14 D-N}. Also the Sun neutrino energy
spectrum has no distortion in the energy region of $E_{\nu_e} =
0.816 \div 13$ MeV., which cannot be in the case if the resonance
mechanism is realized. However some authors insisted and continue
to insist that this mechanism has already been confirmed at
present time.
\par
In the author's works \cite{15 Kh. Besh} two remarks were done:
1)the Wolfenstein's equation is a left-right symmetrical one while
the weak interactions are left-handed interactions (then this
equation has no connection with the weak interactions), 2) Since
the weak interactions with the charged current are the left side
ones, then these interactions cannot generate masses (masses can
be generated only in the left-right symmetric interactions), then
neutrino effective masses cannot change in matter and resonance
conversion will be absent.
\par
Below we show that the problem of resonance enhancement of
neutrino oscillations in matter has a fundamental sense to verify
the weak interactions theory and therefore we must obtain a strict
and direct proof of this effect (the used $\chi^2$ method
\cite{16} is not sufficient for this purpose).

\section{Why is This Resonance Mechanism Important for Verification
of the Weak Interactions Theory?}

It is necessary to remark that changing of couple constants
(running coupling constants) of the strong, electromagnetic and
weak interactions arises due to vacuum polarization.
\par
The weak interactions theory demands a strong and precise check up
and the resonance mechanism gives us these possibilities. The
phenomena are: distortion of the Sun neutrino spectrum, Day-Night
effect and resonance effect in the Earth matter at appropriate
neutrino energies since they are direct consequences of the weak
interactions. In the accelerator experiments with neutrino we
cannot avoid influence of the strong and electromagnetic
interactions in order to separate the contribution of weak
interaction running coupling constant from the contribution of
running coupling constant of these interactions. Since $W, Z^o$
masses are a very big consequently deposit of the weak
interactions running coupling constant in accelerator processes is
very small in comparison with the above mentioned interactions
\cite{17 okun}.
\par
The physics of weak interactions have bad luck from the very
beginning. After discovery of the nuclear beta decay it has been
supposed that in this processes the law of energy momentum
conservation does not fulfil. The situation was corrected after V.
Pauli's letter to the Tubingen Physical Society \cite{18 Pauli}.
Afterwards E. Fermi offered a famous beta decay theory \cite{19
Fermi}. For the first time the interaction generated by neutrino
was observed in experiment of F. Reines and C. L. Cowan  \cite{20
Reines}.
\par
It is necessary to remark that in the standard theory of neutrino
oscillations it is supposed that $\nu_e, \nu_\mu, \nu_\tau$
neutrinos have no definite masses \cite{3}, \cite{21} (definite
masses have only $\nu_1, \nu_2, \nu_3$ neutrinos). Since $\nu_e,
\nu_\mu, \nu_\tau$ neutrinos have no definite masses, then we
cannot formulate the law of energy momentum conservation in the
processes with  neutrino participation. Work \cite{22} suggested a
scheme of neutrino oscillations where the law of energy momentum
conservation is fulfilled.
\par
It is necessary to stress that at neutrino passing through matter
there can be two types of processes (we neglect inelastic
interactions):
\par
\noindent a) elastic scattering and
\par
\noindent b) polarization matter by neutrino.
\par
\noindent With a naive point of view, in analogy with the strong
and electromagnetic interactions, we can suppose that the both
processes will take place at neutrino passing through matter via
weak interactions. It is a usual practice. However, we know that
the weak interactions with the charged current are left-handed
type interactions therefore it is needed to prove that there is
analogy with the above interactions. It is necessary to remark
that if the weak interactions cannot generate masses, then
resonance enhancement of neutrino oscillations in matter can be
realized only at violations of the law of energy momentum
conservation
(see ref. \cite{23} and also below).  \\

\section{Elements of Theory (Mechanism) of Resonance Enhancement
of Neutrino Oscillations in Matter and Some Critical Remarks}

\par
Before consideration of the resonance mechanism it is necessary to
gain an understanding of the physical nature origin of this
mechanism. As stressed above, at neutrino passing through matter
there can be two processes- neutrino scattering and polarization
of the matter by neutrino. Obviously resonance enhancement of
neutrino oscillations in matter will arise due to polarization of
the matter by neutrino. If the weak interaction can generate not
only neutrino scattering but also polarization of matter, then the
resonance effect will exist, otherwise this effect cannot exist.
\par
In the ultrarelativistic limit, the evolution equation for the
neutrino wave function $\nu_{\Phi} $  in matter has the following
form \cite{8}:
$$
i \frac{d\nu_{Ph}}{dt} = ( p\hat I + \frac{ {\hat M}^2}{2p} + \hat
W ) \nu_{Ph} , \eqno(1)
$$
where $p, \hat M^{2}, \hat W_i $ are, respectively, the momentum,
the (nondiagonal) square mass matrix in vacuum, and  the matrix,
taking  into account neutrino interactions in matter,
$$
\nu_{Ph} = \left (\begin{array}{c} \nu_{e}\\
\nu_{\mu} \end{array} \right) , \qquad \hat I = \left(
\begin{array}{cc} 1&0\\0&1 \end{array} \right) ,
$$
$$
\hat M^{2} = \left( \begin{array}{cc} m^{2}_{\nu_{e}\nu_{e}}&
m^{2}_{\nu_{e} \nu_{\mu}}\\ m^{2}_{\nu_{\mu}\nu_{e}}&
m^{2}_{\nu_{\mu} \nu_{\mu}} \end{array} \right).
$$
\par
If we suppose that neutrinos in matter behave analogously to the
photon in matter (i.e., the polarization at neutrino passing
through matter arises) and the neutrino refraction indices are
defined by the expression
$$
n_{i} = 1 + \frac{2 \pi N}{p^{2}} f_{i}(0) = 1 + 2 \frac{\pi
W_i}{p} , \eqno(2)
$$
where $i$ is a type of neutrinos $(e, \mu, \tau)$, $N$ is density
of matter, $f_{i}(0)$ is a real part of the forward scattering
amplitude, then $W_i$ characterizes polarization of matter by
neutrinos (i.e. it is the energy of matter polarization).
\par
The electron neutrino ($\nu_{e}$)  in matter interacts via
$W^{\pm}, Z^{0}$ bosons and $\nu_{\mu}, \nu_{\tau}$ interact only
via $Z^{0}$ boson. These differences in interactions lead to the
following differences in the refraction coefficients of $\nu_{e}$
and $\nu_{\mu}, \nu_{\tau}$
$$
\Delta n = \frac{2 \pi N}{p^{2}} \Delta f(0) , \eqno(3)
$$
$$
\Delta f(0) =  \sqrt{2} \frac{G_F}{2 \pi} p ,
$$
where $G_F$ is the Fermi constant.
\par
Therefore the velocities (or effective masses) of $\nu_{e}$ and
$\nu_{\mu}, \nu_{\tau}$ in matter are different. And at the
suitable density of matter this difference can lead to a resonance
enhancement of neutrino oscillations in matter \cite{8}, \cite{9}
\par
$$
\sin ^{2} 2\theta _{m} = \sin^{2} 2\theta \cdot [(\cos 2\theta  -
{L_{0}\over L^{0}})^{2} + \sin ^{2} 2\theta ]^{-1} , \eqno(4)
$$
where $\sin ^{2} 2\theta _{m}$ and $\sin^{2} 2\theta$ characterize
neutrino mixings in matter and vacuum, $L_{0}$ and $L^{0}$ are
length of oscillations in vacuum and matter
$$
L_{0} = \frac{4 \pi E_{\nu} \hbar}{\Delta m^2 c^3} \qquad L^{0} =
\frac{\sqrt{2} \pi \hbar c}{G_F n_e}, \eqno(5)
$$
where $E_{\nu}$ is neutrino energy, $\Delta m^2$ - difference
between squared neutrino masses, $c$ is light velocity, $\hbar$ is
Plank constant, $G_F$ is fermi constant and $n_e$ is electron
density of matter.
\par
\noindent At resonance

$$ \cos  2\theta  \cong  {L_{0}\over
L^{0}}\qquad sin^{2} 2\theta_{m} \cong 1\qquad \theta_{m} \cong
{\pi \over 4} . \eqno(6)
$$
\par
It is necessary to stress that this resonance enhancement of
neutrino oscillation in matter is realized when neutrino velocity
is less than the light velocity in matter (i.e. $ v_i <
\frac{c}{n_i})$.

\par
As we can see from the form of Eq. (1), this equation holds the
left-right symmetric neutrinos wave function $\Psi(x) = \Psi_L(x)
+ \Psi_R(x)$. This equation contains the term $W$, which arises
from the weak interaction (contribution of $W$ boson) and which
contains only a left-handed interaction of the neutrinos, and is
substituted in the left-right symmetric equation (1) without
indication of its left-handed origin. Then we see that equation
(1) is an equation that includes term $W$ which arises not from
the weak interaction but from a hypothetical left-right symmetric
interaction (see  also works \cite{24}, \cite{25}, \cite{26 besh}.
Therefore this equation is not the one for neutrinos passing
through real matter. The problem of neutrinos passing through real
matter has been considered in \cite{24}, \cite{25}, \cite{26
besh}, \cite{23}.
\par
Then there is a question: Can this resonance effect exist if the
weak interactions do not generate masses (i.e. do not change
neutrino masses in matter)?

\par
\section{What problem arises in Weak Interactions Theory
in Contrast to the Strong and Electromagnetic Interactions
Theory?}
\par
In strong and electromagnetic interactions the left-handed and
right-handed components of spinors participate in interactions in
symmetric manner. In contrast to these interactions only the
left-handed components of spinors participate in the weak
interactions with charged current (it is also necessary to remark
that in the weak neutral current the left-handed and right-handed
components of spinors participate in non symmetric manner). This
is a distinctive feature of the weak interactions.

\par
\noindent
\subsection{Elements of the Electroweak Interactions
Model}

\par
Electroweak interaction lagrangian includes the following lepton
and quark doublets
\par
$$
\Psi _{lL} = \left(\begin{array}{c} \nu _{l}\\
l\end{array}\right)_L ,
 \Psi_{lR} , l = e, \mu , \tau
$$
\vspace{1.0cm}
$$
\hspace{1cm}i = 1 \hspace{1cm} i = 2 \hspace{1cm} i = 3
$$
$$
\Psi _{iL} = \left(\begin{array}{c} u\\ d \end{array} \right)_L ,
\left(\begin{array}{c} c\\ s \end{array} \right)_L ,
\left(\begin{array}{c} t\\ b \end{array} \right)_L , \eqno(7)
$$
and left components of charged leptons and quarks
$$
\Psi _{iR} = u_{R}, d_{R};\qquad c_{R}, s_{R};\qquad t_{R}, b_{R}
.
$$
\par
And this lagrangian has the following form \cite{27 WSG}:
$$
{\cal L}_{I} = ig j^{K,\alpha } A^{K}_{\alpha}  + ig'\frac{1}{2}
j^{\Upsilon ,\alpha } B_{\alpha } , \eqno(8)
$$
where
$$
j^{K,\alpha } = \sum^{3}_{i=1} \bar \Psi _{i,L} \gamma ^{\alpha }
{\tau^{K}\over 2} \Psi _{i,L} +
$$
$$
\sum_{l=e,\mu,\tau}  \bar \Psi_{l,L} \gamma ^{\alpha }
{\tau^{K}\over 2} \Psi _{l,L} , \eqno(9)
$$
and
$$
{1\over 2} j^{\Upsilon,\alpha } = j^{em,\alpha } - j^{3,\alpha } ,
$$
($j^{em,\alpha }$- electromagnetic current of quarks and leptons),
where $A^{i}_{\alpha }, B_{\alpha }$- are gauge fields associated
with $SU(2)_{L}$  и $U(1)$ - groups; $\Upsilon $ - hypercharges of
quarks and leptons.
\par
At transition from $A^{3}_{\alpha }, B_{\alpha }$ fields to
$Z_{\alpha }, A_{\alpha }$ fields
$$
Z_{\alpha } = A^{3}_{\alpha } \cos  \theta _{W} - B_{\alpha } \sin
\theta _{W} , \eqno(10)
$$
$$
A_{\alpha } = A^{3}_{\alpha } \sin  \theta _{W} + B_{\alpha } \cos
\theta _{W} ,
$$
interaction lagrangian for $Z_{\alpha }, A_{\alpha }$ fields gets
the following form:
$$
{\cal L}^{o}_{I} = i {g\over 2 \cos  \theta _{W}} j^{o,\alpha }
Z_{\alpha }
 + ie j^{em,\alpha } A_{\alpha } ,
\eqno(11)
$$
where $j^{o,\alpha } = 2 j^{3,\alpha } - 2 \sin ^{2} \theta _{W}
 j^{em,\alpha }$ - is neutral current of the standard model.

\par
\noindent
\subsection{Running Coupling Constant in the
Standard Weak Interactions Model}

\par
It is supposed that in the electroweak model  the coupling
constants $g, g'$ depend on transfer momenta \cite{28} and
equation for $g(Q^2)$ has the following form:
$$
\frac{d g^{-1}}{d (ln Q^2)} = \frac{1}{4 \pi} \left[ \frac{22}{3}
- \frac{4 F}{3} \right] , \eqno(12)
$$
where $F$ is family numbers ($F = 3$) (here we consider only a
weak part of the electroweak model since in the electromagnetic
interactions there is renormalization of the coupling constant).
It means that in the weak interactions the vacuum polarization
takes place as in the strong and electromagnetic interactions. It
is necessary to remember that in the weak interactions in contrast
to these interactions only the left components of fermion
participate in the weak interactions of charged current (as
stressed above in the weak neutral current the left-handed and
right-handed components of spinors also participate in non
symmetric manner).
\par
If the coupling constant of the weak interaction  is renormalized
then the effective masses of fermions in matter also change, i.e.
the standard weak interaction can generate effective masses. It
means that at the weak interactions the resonance enhancement of
neutrino oscillations in matter will take place
\cite{8}-\cite{10}. It is necessary to keep in mind that our
consideration refers only to the weak interaction with charged
current.

\par
\noindent
\subsection{Remarks About the Coupling Constant of the
Standard Weak Interactions Model}

\par
As we have stressed above, a distinctive feature of the weak
interactions is violation of $P$ parity. Now let us consider the
consequences of the distinctive feature for coupling constant of
the weak interactions.

\par
The simplest method to prove the absence of the polarization in
vacuum and matter is \cite{29}:
\par
   If we put an electrical (or strong) charged particle in  vacuum,
polarization of vacuum will appear. Since the field around the
particle is spherically symmetrical, the polarization must be also
spherically symmetrical. Then the particle will be left at rest
and the law of energy and momentum conservation is fulfilled.
\par
If we put a weakly interacting particle (a neutrino) in vacuum
then, since the field around the particle has a left-right
asymmetry (weak interactions are left interactions with respect to
the spin direction), polarization of vacuum must be nonsymmetric,
i.e., on the left side there will be maximal polarization and on
the right side there will be zero polarization. Since polarization
of the vacuum is asymmetrical, there arises an asymmetrical
interaction of the particle (the neutrino) with vacuum and the
particle cannot be at rest and will be accelerated. Then neutrino
will get energy-momentum from the vacuum and the law of energy
momentum conservation will be violated. The only way to fulfil the
law of energy-momentum conservation is to require that
polarization of vacuum be absent at the weak interactions. The
same situation will take place in matter (do not mix up it with
particle acceleration at the weak interactions!) .
\par
It is interesting to remark that in the gravitational interaction
the polarization does not exist either \cite{30}. \\

\subsection{Is Mass Generation Possible in Weak Interactions?}

\par
It is well known that masses are generated in the strong and
electromagnetic interactions. Is mass generation possible in the
weak interactions? This question arises for the left-handed
character charged current of the weak interactions. Let us
consider consequences of this feature.
\par
We will show that the lepton masses cannot be generated in the
framework of the electroweak interactions model \cite{26 besh} or
in the weak interactions model included in this electroweak
interactions model as a component. Consideration will be carried
out for $U(1)$ theory and then it will be generalized for $SU(2)$
theory.
\par
Dirac equation for the lepton (spinor) wave function  $\psi = \psi
_{R} + \psi _{L}$ in external field $A_{\mu }$ has the following
form:
\par
$$
( E + \sigma _{i}H_{i}) \psi _{L} - M \psi _{R} = 0 , i=1-3 ,
\eqno(13)
$$
$$
( E - \sigma _{i}H_{i}) \psi _{R} - M \psi _{L} = 0 , E= \epsilon
-e A_{4} ,
$$
where $H_{i} = P_{i} - e A_{i},\qquad  \sigma _{i}$ - Pauli
matrices.
\par
The same equation without external field $A_{\mu }$ can be written
in the following form:
\par
$$
( E' + \sigma _{i}P'_{i}) \psi _{L} - M' \psi _{R} = 0 ,\eqno(14)
$$
$$
( E' - \sigma _{i}P'_{i}) \psi _{R} - M' \psi _{L} = 0 .
$$
From  (13) и (14) and using that $\Delta M = M - M'$ we obtain
\par
$$
( (E - E') + \sigma _{i}(H_{i}- P'_{i}) ) \psi _{L} = \Delta M
\psi _{R} , \eqno(15)
$$
$$
( (E - E') - \sigma _{i}(H_{i}- P'_{i}) ) \psi _{R} = \Delta M
\psi _{L} ,
$$
that deposit of the interaction caused by the external field
$A_{\mu }$ leads to appearance of masses difference $\Delta M$
which is symmetric to the left and right components of fermion.
Now using expression (15) we can consider the case when in
interaction there is only a left component of spinor (fermion) as
it takes place in the weak interactions. Then expression (15) can
be rewritten in the form:
$$
( (E - E') + \sigma _{i}(H_{i}- P'_{i}) ) \psi _{L} = 0 ,
\eqno(16)
$$
$$
\hspace{6.cm}0 = \Delta M \psi _{L} ,
$$
if in (16) $\psi _{R} = 0$ and $\psi _{L}$ differs from zero, then
$\Delta M = 0.$
\par
So, if only the left-handed component of fermion participates in
the interaction, then the fermion mass does not change. It is not
difficult to generalize the above considered case $U(1)$ theory to
the case of $SU(2)$ theory and then we come to a conclusion that
since the right-handed components of fermions do not participate
in the weak interactions, then these interactions cannot generate
masses.
\par
For obtaining the masses in the standard model of electroweak
interactions founded on group $SU(2)_{L} \times U(1)$  the Higgs
mechanism \cite{31 higgs} is used.
\par
So, we come to a conclusion that the standard weak interactions by
charged current cannot generate masses for their left-handed
character.
\par
If we consider neutrino oscillations in the scheme of mass mixings
\cite{21}, \cite{28}
$$
M = \left(\begin{array}{cc}m_{\nu_e} & m_{\nu_e \nu_\mu} \\
m_{\nu_\mu \nu_e} & m_{\nu_\mu} \end{array} \right) , \eqno(17)
$$
where $m_{\nu_e}, m_{\nu_\mu}$ are masses of $\nu_e, \nu_\mu$
neutrinos and $m_{\nu_e \nu_\mu}, m_{\nu_\mu \nu_e}$ are
nondiagonal mass terms, then the expression for $sin^2 (2 \theta)$
has the following form:
$$
sin^2 (2\theta) = \frac{(2m_{\nu_{e} \nu_{\mu}})^2} {(m_{\nu_e} -
m_{\nu_\mu})^2 +(2m_{\nu_e \nu_{\mu}})^2} . \eqno(18)
$$
Since the weak interactions cannot generate masses, then masses
$m^{matt}_{\nu_e}$, $m^{matt}_{\nu_\mu}$ of $\nu_e, \nu_\mu$
neutrinos in matter do not change
$$
m^{matt}_{\nu_e} = m_{\nu_e}, \quad m^{matt}_{\nu_\mu} =
m_{\nu_\mu} ,
$$
and then
$$
sin^2_{matt} (2\theta) = sin^2 (2\theta) \eqno(19),
$$
i.e., the mixing angle $\theta$ does not change in matter and
resonance effect must not exist.

\subsection{How Can this Resonance Effect  be Realized if the Weak
Interactions do not Generate Masses?}

\par
The law of energy momentum conservation of particle (neutrino) has
the following form (we take only matter polarization into account)
\cite{32}:
$$
a) \hspace {0.2cm} E_ {0} = E + W,
$$
$$
b) \hspace {0.2cm} p_ {0} = p + W \beta, \eqno (20)
$$
where $E_ {0}, p_ {0}, E, p$, are correspondingly, energy and
momentum of neutrino in vacuum and matter, $W$ - energy
polarization matter by neutrino, $c = 1 \quad \beta = \frac {v}
{c} \to v, \quad \frac{W}{c^2} \to W$.
\par
It is obvious that neutrino matter polarization (reaction) moves
in matter with the velocity $\beta$ which is equal to neutrino
velocity.
\par
If to put expression b) in expression a), then we obtain the
following:
$$
\sqrt {p_ {0} ^2 + m_ {0} ^2} = \sqrt {(p_{0} - W \beta) ^ {2} +
m_{0}^ {2}} + W \to
$$
$$
W[(1 - \beta^2) W - 2 \sqrt {p_ {0} ^2 + m_ {0} ^2} + 2 p_{0}
\beta] = 0 , \eqno (21)
$$
what has a solution only if (obviously, if $m_0 = 0$ then $\beta =
1$ ):
$$
m_{0} = 0,
$$
or
$$
W = 0. \eqno (22)
$$
The demand of fulfilment of the law of energy momentum
conservation for the weak interacting particle (neutrino) in
matter will result in conclusion that if
\par
$ m_{0} = 0$,\qquad then \qquad $W \neq 0$,
\par
and if
\par
$m_{0} \neq 0$, \qquad then \qquad $W = 0$ . \hspace{6.5cm} (23)
\par
\noindent These decisions mean that for massive neutrino the
matter polarization energy $W$ must be zero, i.e., $W = 0$ in
contrast to energy polarization charged particle in
electrodynamics.
\par
Since Wolfenstein's equation describes massive neutrinos, then $W$
must be equal to zero. Therefore here an enhancement of neutrino
oscillation in matter cannot arise.
\par
Otherwise when $m_ {0} = 0$ the polarization energy $W$ can differ
from zero, but as it is well known that massless neutrinos cannot
oscillate. The intermediate case when some neutrinos have masses
and the rest neutrinos are massless, is of no interest since it is
clear how neutrinos which participating in the weak interactions
(having a weak charge) can be massless.
\par
So, we come to a conclusion that, if we demand fulfillment of the
law of energy momentum conservation in matter, then the resonance
enhancement of neutrino oscillations cannot appear. And it is a
right consequence of the left-handed character of the weak
interactions by charged current.
\par

\section{What is the Situation with Experimental Confirmation of this
Resonance Mechanism?}

\par
At present the experimental data have been obtained on the
accelerator, reactor, atmospheric and solar neutrinos. The data
obtained in the reactor, accelerator and atmospheric neutrinos
have shown that the $\theta_{1 2}, \theta_{2 3}$ have big values.
The estimation of the value of this angle can be extracted from
KamLAND \cite{33} data and it is:
$$
sin^2 (2 \theta_{1 2}) \cong 1.0 , \quad \theta \cong
\frac{\pi}{4}, \quad \Delta m^2_{1 2} = 6.9 \cdot 10^{-5} {eV}^2,
\eqno(24)
$$
or
$$
sin^2 (2 \theta_{1 2}) \cong 0.83 , \quad \theta_{1 2} = 32^o ,
\quad \Delta m^2_{1 2} = 8.3 \cdot 10^{-5} {eV}^2.
$$
The angle mixing for vacuum $\nu_\mu \to \nu_\tau$ transitions
obtained on SuperKamiokande \cite{34} for atmospheric neutrinos
is:
$$
sin^2 (2\gamma_{2 3})  \cong  1, \quad \gamma \cong \frac{\pi}{4}
\quad \Delta m^2_{2 3} \simeq 2.5 \cdot 10^{-3} {eV}^2. \eqno(25)
$$
\par
The value of the Solar neutrinos flow measured (through elastic
scattering) on SNO \cite{35} is in a good agreement with the same
value measured in SuperKamiokande \cite{36}.
\par
Ratio of $\nu_e$ flow measured on SNO (CC) to the same flow
computed in the frame work of SSM \cite{37} ($E_\nu > 6.0 MeV$)
is:
$$
\frac{\phi_{SNO}^{CC}}{\phi_{SSM2000}} = 0.306 \pm 0.026(stat.)\pm
0.024(syst.) . \eqno(26)
$$
This value is in a good agreement with the same value of $\nu_e$
relative neutrinos flow measured on Homestake (CC) \cite{38} for
energy threshold $E_\nu = 0,814 MeV$
$$
 \frac{\Phi^{exp}}{\Phi^{SSM2000}} = 0.34 \pm 0.03 .
\eqno(27)
$$
From these data we can come to a conclusion that the angle mixing
for the Sun $\nu_e$ neutrinos does not depend on neutrino energy
thresholds (0.8 $\div$ 13 MeV) and in this region the energy
spectrum has no distortion.

\begin{center}
\mbox{\epsffile{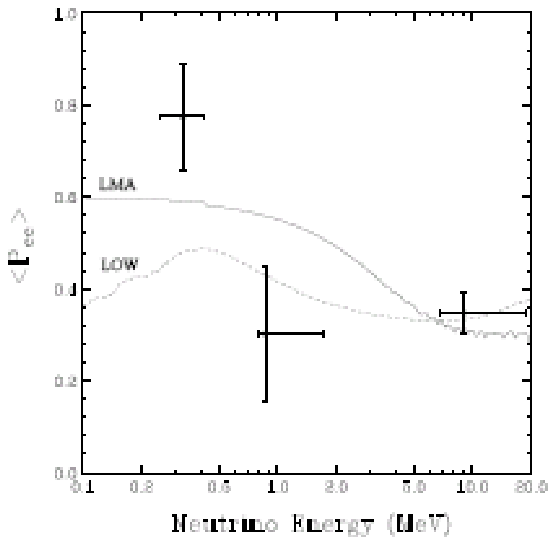}} \vspace{2mm} \noindent
\end{center}

{\sf Figure~1:}~  The profile of the effect. Shown are the
reconstructed values of the survival probability in different
energy ranges. The lines correspond to the survival probability
for the LMA and LOW solutions; (from \cite{41}). \\

\par
The survival probability in different energy ranges of the solar
neutrinos \cite{39} (see also ref. \cite{40 Kayser}) was computed
taking into account the resonance effect. The profile of this
effect is shown in Figure 1 (shown are the reconstructed values of
the survival probability in different energy ranges. The lines
correspond to the survival probability for the LMA and LOW
solutions (from \cite{41}).
\par
From the above Figure 1 we see that the curves obtained from the
computation in the framework of the resonance mechanism \cite{39}
are in clear discrepancy with the above given experimental data
(also see below Figure 5). In spite of this fact some authors come
to a conclusion that this mechanism has been proved in experiments
(experimental errors given in this figure many times exceed the
same published errors, it is necessary to suppose that these
errors were smeared for obtaining small values for $\chi^2$ or
better adjustment at smaller value of $\sigma$). The same
situation takes place in the last interpretations of the solar
neutrino data \cite{16}, \cite{42}. The energy profile of the
solar $E_\nu$ survival probability $P_{ee}$ for best-fit LMA
values ($\theta_{1 3} = 0$) is shown in Figure 2 (experimental
data see in Figure 4 and 5 also in expressions (24)-(27)). Value
for $\theta_{1 3} = 0$ was obtained from CHOOZ result analysis
\cite{chooz}\\

\par
\noindent {\bf Is the CHOOZ result analysis trustful (i.e., is it
correct that $\theta_{1 3} = 0$)?}
\par
The probability of $P_{\bar \nu_e \bar \nu_e}$ transitions at
three neutrino oscillations is:
$$
P_{\bar \nu_e \to \bar \nu_e} (R)= 1 - cos^4(\theta_{1 3})sin^2(2
\theta_{1 2}) sin^2(\frac{R}{L_{1 2}}) -
$$
$$
cos^2(\theta_{1 2}) sin^2(2 \theta_{1 3}) sin^2(\frac{R}{L_{1 3}})
-sin^2(\theta_{1 2}) sin^2(2 \theta_{1 3}) sin^2(\frac{R}{L_{2
3}})  \eqno(28),
$$
where $L_{1 2}, \quad L_{1 3} \quad L_{2 3}, \quad R$ ,
correspondingly, are lengths of neutrino oscillations and distance
from neutrino source. Since $\quad L_{1 3} \approx L_{2 3}$, we
can rewrite expression (28) in the following form:
$$
P_{\bar \nu_e \to \bar \nu_e} (R) \approx 1 - cos^4(\theta_{1
3})sin^2(2 \theta_{1 2}) sin^2(\frac{R}{L_{1 2}})- sin^2(2
\theta_{1 3}) sin^2(\frac{R}{L_{1 3}}) , \eqno(29)
$$
if $L_{1 2} >> R$, and taking into account that $\frac{L{1
2}}{L_{2 3}} \approx 30.5$, then the above expression can be
rewritten in the following form:
$$
P_{\bar \nu_e \to \bar \nu_e} (R) \approx 1 - sin^2(2 \theta_{1
3}) sin^2(\frac{R}{L_{1 3}}) , \eqno(30)
$$


\par
\begin{center}
\mbox{\epsffile{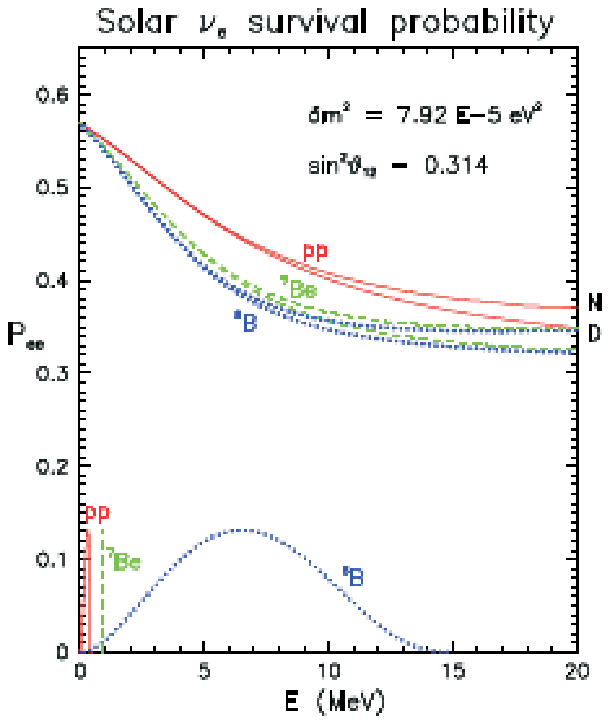}} \vspace{2mm} \noindent
\end{center}

Figure 2: The energy profile of the solar $E_\nu$ survival
probability $P_{ee}$ for best-fit LMA values and $\theta_{1 3} =
0$. The function $P_{e e}($E) shows a smooth transition from
vacuum to the matter dominated regime as E increases, with some
differences induced by averaging over different production regions
(for 8B, 7Be and pp neutrinos) and, to a smaller extent, by
nighttime (N) Earth effects with respect to daytime (D). Also
shown are the corresponding solar neutrinos
energy spectra (in arbitrary vertical scale). \\

\par
\noindent since $L_{1 2} \approx 160$ km, $R_{CHOOZ} \approx 1$
km, then $\frac{R}{L_{1 3}} \approx 5.3, \quad sin^2(\frac{R}{L_{1
3}}) \approx \frac{1}{28} = 0.036 $. The expression for transition
probability $ P_{\nu_e \to \nu_e} (R_{CHOOZ})$ is
$$
P_{\bar \nu_e \to \bar \nu_e} (R_{CHOOZ}) \approx 1 - 0.036 \cdot
sin^2(2 \theta_{1 3})  , \eqno(31)
$$
and then the value of $1 - P_{\bar \nu_e \to \bar \nu_e}
(R_{CHOOZ})$ cannot be larger than $0.036$:
$$
1 - P_{\bar \nu_e \to \bar \nu_e} (R_{CHOOZ}) \leq 0.036 .
$$
The precision of the CHOOZ experiment is $\approx 5 \%$, i.e.
0.05. It is clear that for obtaining a limitation on $sin^2(2
\theta_{1 3})$ the precision of this experiment must be less than
0.036. So, we see that in this type of experiment a proper
limitation on $sin^2(2 \theta_{1 3})$ is possible to obtain only
if distances $R$ are $3 \div 5$ km or if the precision of the
experiment is very big ($\approx 0.4\div 0.5 \%$).
\par
Now there is a new mechanism of enhancement of neutrino
oscillation which \cite{barger43} is named as MaVaN (mass-varying
neutrino oscillations) mechanism. The result of computation in the
framework of this mechanism together with the profile of the MSW
effect is given in Figure 3. We will not discuss this mechanism
since at present a direct confirm of the dark matter existence is
absent as well as its weak interactions with neutrinos.
\vspace{1cm}

\par
\begin{center}
\mbox{\epsffile{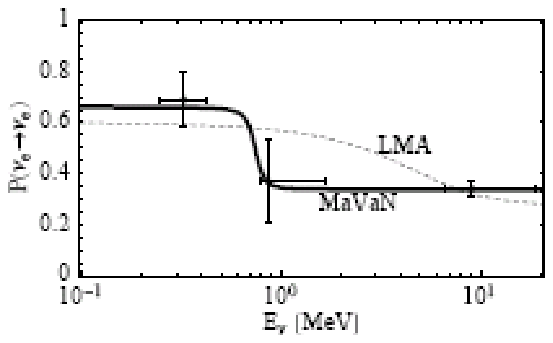}} \vspace{2mm} \noindent
\end{center}

{\sf Figure~3:}~ $P(\mu_e \to \nu_e)$  vs. $E_\nu$ for MaVaN
\cite{barger43} oscillations (solid curve). The dashed curve
corresponds to conventional oscillations with the best-fit
solution to KamLAND data.

\vspace{1cm}

\par
Figure 5 gives the profile of the MSW effect (i.e. the
reconstructed values of the survival probability in different
energy ranges for the LMA solution from [41]). The following
experimental data are also shown:
\par
\noindent 1. From the Homestake experiment in 1970-1994y.
\cite{38} where the relation between the measured and calculated
\cite{37} flux data is
$$
\frac{\Phi^{exp}}{\Phi^{SSM2000}} = 0,34 \pm 0,03 , \eqno(32)
$$
\par
\noindent 2. From the GALLEX (GNO) \cite{bellotti44}, \cite{c46}
and SAGE \cite{gavrin45}, \cite{c46} experiments where the
relation between measured and calculated BP04 \cite{b-2004 48}
flux data are
$$
 \frac{\Phi^{exp}_{GALLEX}}{\Phi^{BP04}} = 0,53 \pm 0,04 ,
\eqno(33)
$$
$$
 \frac{\Phi^{exp}_{SAGE}}{\Phi^{BP04}} = 0,51 \pm 0,04 .
\eqno(34)
$$
The data from Ga-Ge experiments are placed higher than the data of
other experiments. It is necessary especially to note that the
value of these experimental data decreases with statistics
increasing.
\par
\noindent 3. From the SNO \cite{35} experiment where the relation
between the measured and calculated SSM2000 \cite{37} flux data
are
$$
\frac{\phi_{SNO}^{CC}}{\phi_{SSM2000}} = 0,35 \pm 0,02 , \eqno(35)
$$
and \cite{b-2000 49}
$$
\frac{\phi_{SNO}^{CC}}{\phi_{SSM2000}} = 0,309 \pm 0,02 ,
\eqno(36)
$$
\par
\noindent 4. From the SuperKamiokande \cite{36} experiment where
the relation between the measured and calculated SSM2000 \cite{37}
flux data is
$$
\frac{\Phi^{total}_{^{8}B}}{SSM2000} = 0.465 \pm 0,005(stat) +
0.016 (-0.015)(syst), \eqno(37)
$$
The data in Figure 5 above 5 MeV were obtained by subtraction of
the neutral current ($Z^o$ boson) deposit obtained in SNO from the
SuperKamiokande data (see Figure 4) and this difference equals to
$\Delta = 0.156$ (it is the difference between the values of $
\frac{\Phi^{total}_{^{8}B}}{SSM2000}$ in exp. (37) and $
\frac{\phi_{SNO}^{CC}}{\phi_{SSM2000}}$ in exp. (36)). The
theoretical value of $\Delta$ is $\Delta \approx 0.155$.
\par
From Figure 5 we see that the data obtained in SuperKamoikande,
Homestake do not coincide with the computation obtained on the
resonance effect in matter, i.e., the resonance effect is not
confirmed. Only one point obtained in GALLEX and SAGE comes out
from the other neutrino experimental data. Therefore it is very
important to study the solar neutrinos energy spectrum below 1 MeV
to clarify the reason of this deviation.

\begin{figure}[h!]
   \begin{center}
   \includegraphics[width=8cm]{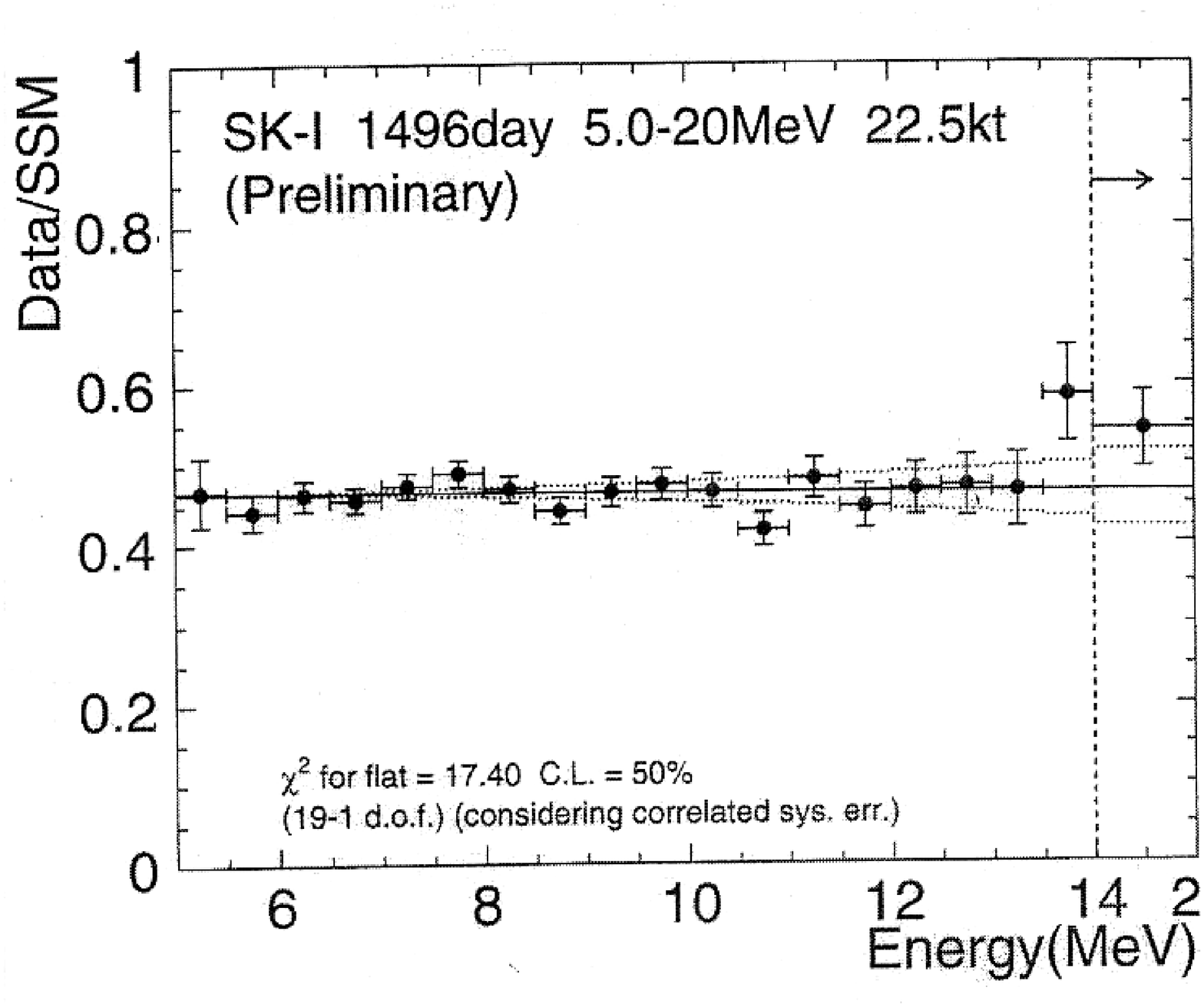}
   \end{center}
   \end{figure}

{\sf Figure~4:}~The energy profile of the solar $E_{\nu_e}$
neutrinos flux from SuperKamoikande experiment ($P_{\nu_e}
(E_\nu)/ P_{SSM2000}(E_\nu))$. \\

\par
 The Day-Night effect is not confirmed. Usually it is claimed that this
effect a very small. To avoid this argumentation it is necessary
to carry out an experiment with the bigger statistics (for
example, in SuperKamiokande). This problem also can be solved by
using neutrinos passed through the Earth at resonance energies for
the Earth densities
$$
E_{res} = \frac{|\Delta m^2| cos 2 \theta_V}{2 \sqrt{2} G_F n_{e,
earth}}, \eqno(38)
$$
where $\theta_V$ is the vacuum angle mixing, $G_F$ is Fermi
constant, $n_{e, earth}$ is electron density of the Earth. \\

\newpage
\begin{figure}[h!]
  \begin{center}
   \includegraphics[width=16cm]{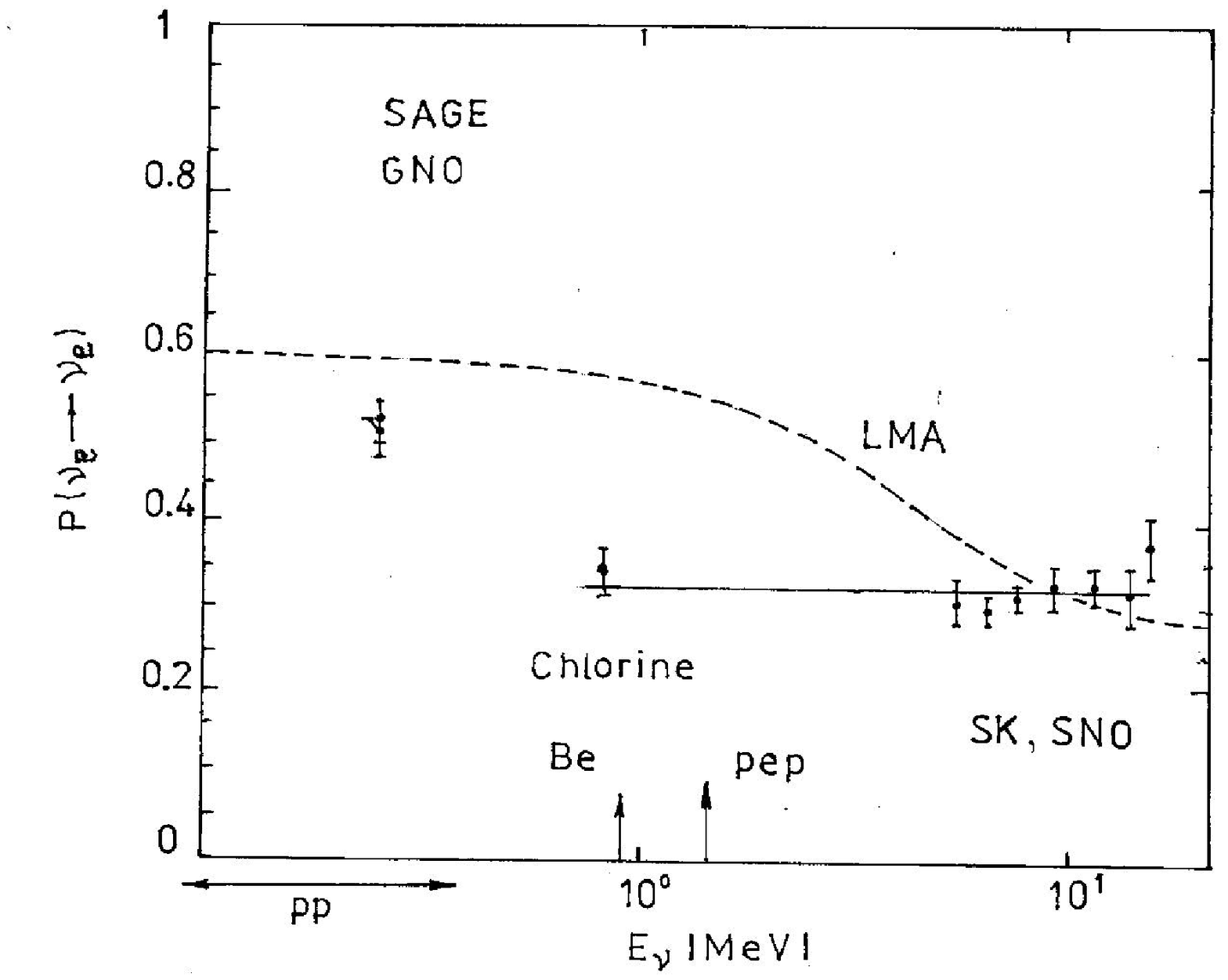}
   \end{center}
   \end{figure}

 {\sf Figure~5:}~The energy profile of the solar $E_{\nu}$ survival
 probability $P_{\nu_e \nu_e}$. The point and circles are SAGE, GNO, Chlorine,
 SNO and SuperKamoi-kande experimental data. The dashed curve corresponds
to the profile of MSW effect [39].

\newpage
\section{Conclusion}

\par
The mechanism of resonance enhancement of neutrino oscillations in
matter and some critical remarks to this mechanism have been
considered. Wolfenstein's equation (1) contains term $W$, which
arises from the weak interaction (contribution of $W$ boson) which
is the only left-handed interaction of the neutrinos, and it is
substituted in the left-right symmetric equation (1) without
indication of its left-handed origin. Then we see that equation
(1) is an equation that includes term $W$ which arises not from
the weak interaction but from a hypothetical left-right symmetric
interaction. Therefore this equation is not the one for neutrinos
passing through the real matter.
\par
Using this resonance mechanism is very important to check the weak
interaction part of the model of electroweak interactions since
the processes induced by this mechanism grow multiply.
\par
In contrast to the electromagnetic and strong interactions in the
weak interactions, the $P$-parity is violated, therefore the
problem of mass generations and charge renormalization in the weak
interactions were considered (the interaction must be left-right
symmetric for mass generations). It is concluded that the
possibility of mass generation and charge renormalization in the
weak interactions has been not proved.
\par
The present status of this resonance effect by using the existence
experimental data has been considered and it is concluded that
this effect has no clear experimental confirmation. For this
purpose it is necessary to fulfil precision experiments with solar
neutrinos and the neutrinos which have passed through the Earth
matter.
 \\

\end{document}